\documentclass[12pt]{iopart}
\usepackage{graphics}
\usepackage{color}

\begin{document}
\title{In situ observation of stress relaxation in epitaxial graphene}
\author{
Alpha T. N'Diaye$^1$\footnote{Corresponding autor; Email: ndiaye@ph2.uni-koeln.de},
Raoul van Gastel$^2$,
Antonio J. Mart\'inez-Galera$^3$,
Johann Coraux$^1$\footnote{Permanent address: Institut N{\'e}el/CNRS-UJF, 25 rue des Martyrs, BP 166, 38042 Grenoble cedex 9, France}, 
Hichem Hattab$^4$,
Dirk Wall$^4$,
Frank-J. Meyer zu Heringdorf$^4$,
Michael Horn--von Hoegen$^4$,
Jos\'e M. G\'omez-Rodr\'iguez$^3$,
Bene Poelsema$^2$,
Carsten Busse$^1$,
Thomas Michely$^1$
}
\address{
{$^1$ {II. Physikalisches Institut}, Universit{\"a}t zu K{\"o}ln, Z{\"u}lpicher Stra{\ss}e 77, 50937 K{\"o}ln, Germany}
{$^2$ MESA+ Institute for Nanotechnology, University of Twente, P.O.Box 217, 7500 AE Enschede, The Netherlands}
{$^3$ Departamento de F\'{\i}sica de la Materia Condensada, C-III, Universidad Aut\'{o}noma de Madrid, E-28049-Madrid, Spain}
{$^4$ Institut f{\"u}r Experimentelle Physik, Universit{\"a}t Duisburg--Essen, Lotharstrasse 1, 47057 Duisburg, Germany}
}

%
\begin{abstract}
Upon cooling, branched line defects develop in epitaxial graphene grown at high temperature on Pt(111) and Ir(111).
Using atomically resolved scanning tunneling microscopy we demonstrate that these defects are wrinkles in the graphene layer, i.e. stripes of partially delaminated graphene. 
With low energy electron microscopy (LEEM) we investigate the wrinkling phenomenon in situ. Upon temperature cycling we observe hysteresis in the appearance and disappearance of the wrinkles. Simultaneously with wrinkle formation a change in bright field imaging intensity of adjacent areas and a shift in the moir\'e spot positions for micro diffraction of such areas takes place. The stress relieved by wrinkle formation results from the mismatch in thermal expansion coefficients of graphene and the substrate. 
A simple one-dimensional model taking into account the energies related to strain, delamination and bending of graphene is in qualitative agreement with our observations. 
\end{abstract}
%
\section{Introduction}
The new material graphene receives currently an enormous attention for its exciting properties.
At the heart of the scientific interest are the consequences of graphene's unique band structure arising from its lattice symmetry and its monoatomic thickness \cite{wallace1947}.
The high mobility of electrons in graphene and the strong electric field effect foster work to realize graphene based electronics \cite{castroneto2009}.
Moreover, the use of graphene for conducting transparent electrodes \cite{reina2009, kim2009}, to realize photosensitive transistors \cite{xia2009}, ultracapacitors \cite{stoller2008}, or a new class of catalytic and magnetic materials through templated cluster growth has been suggested \cite{ndiaye2006, ndiaye2009cluster}.

Although the exciting electronic properties of graphene have been explored mainly by transport measurements for devices built on flakes of exfoliated graphene on SiO$_2$, there appears to be consensus that for future scientific exploration and technological applications epitaxial growth of high quality graphene over large areas \cite{tromp2009, emtsev2009, sutter2008, loginova2008, coraux2009, yu2008, marchini2007} is a prerequisite. 

For any technological application it is of utmost importance to avoid or at least control the defects in graphene associated with the epitaxial growth process.
Such defects may result from growth obstacles caused by substrate steps \cite{emtsev2009,sutter2008} or by the coalescence of finite sized graphene domains \cite{coraux2009}. 

Branched line defects are present in mono- or multilayers of continuous graphene at room temperature after high temperature ($>$1000\,K) epitaxial growth on several metals and on SiC. 
Their nature has been under debate. While some authors attributed the branched line defects to carbon nanotube formation \cite{derycke2002, starr2006, fujita2004, nyapshaev2009}, current research evidences that they are wrinkles in the graphene layers on SiC \cite{cambaz2008,biedermann2009,sun2009,kim2009} as well as on metals \cite{obraztsov2007,chae2009,loginova2009a}. For wrinkles in graphene on SiC very recent atomic resolution  scanning tunneling microscopy (STM) data provide unambiguous evidence for the wrinkle interpretation \cite{sun2009}. Very recent LEEM and STM data provide evidence for the presence of wrinkles also in monolayer graphene on Ir(111) \cite{loginova2009a}. Wrinkle features have also been observed in layered transition metal dichalcogenides \cite{spiecker2006}.
Among other suggestions several authors attribute the occurrence of wrinkles to the difference in the thermal expansion coefficients of graphene and its support \cite{obraztsov2007,cambaz2008,chae2009,sun2009,loginova2009a}. Consistent with that mismatch is the observation of compressive strain in epitaxial graphene at room temperature \cite{hass2008,ferralis2008,ndiaye2006}. This compression was found to vary on a length scale of less then 300\,nm \cite{robinson2009}. Although in situ investigations could provide deeper insight into the wrinkling phenomenon, such investigations are missing till now.

Here we not only support the wrinkle interpretation of the branched line defects by atomically resolved scanning tunneling microscopy (STM) imaging of wrinkles on metals, but also gain a detailed insight into wrinkle formation by in situ LEEM imaging and micro diffraction. Most noteworthy, LEEM and micro diffraction "see" not only wrinkle formation but also the structural and electronic changes in adjacent $\mu$m-sized areas within the graphene. A model is developed, which qualitatively agrees with our observations.

\section{Methods}
We examined epitaxial graphene on Ir(111) and Pt(111). 
Graphene has been grown epitaxially by chemical vapor deposition of ethene (C$_2$H$_4$) at elevated temperatures in ultra high vacuum.
Scanning tunneling microscopy (STM) was carried out at room temperature, low energy electron microscopy (LEEM) imaging was done at variable temperature. 
Growth and imaging was performed in ultra high vacuum without any transfer outside the vacuum.

\section{Results and discussion}
\subsection{Wrinkle formation}
\begin{figure}
\includegraphics{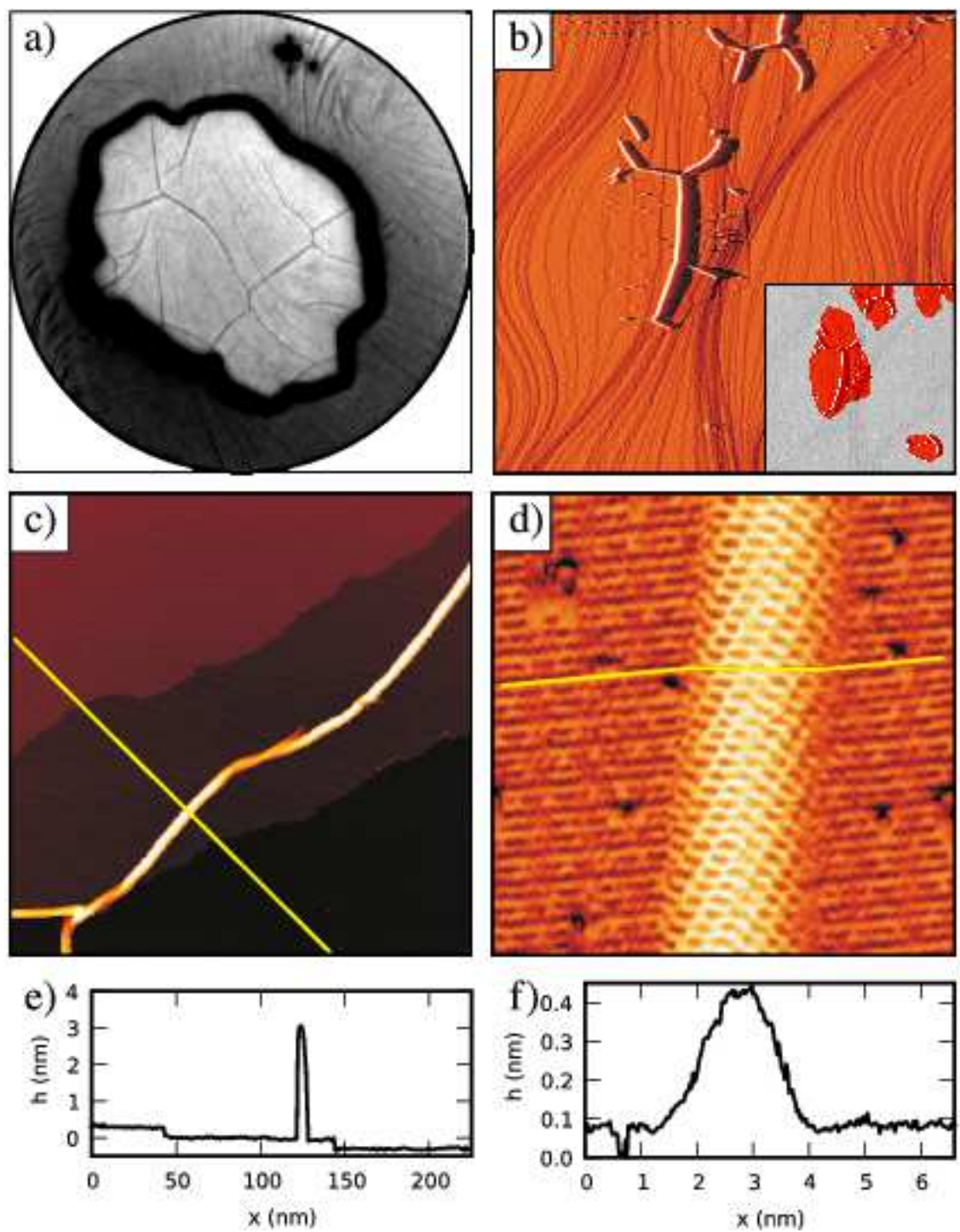}
\caption{
a) LEEM image (field of view: 10\,$\mu$m, electron energy: 2.8\,eV) of a graphene flake on Ir(111) at room temperature.
b) STM topograph of graphene flakes on Pt(111) ($3\,\mu m \times 3\,\mu m$). The image is differentiated and appears as if illuminated from the left. 
The inset shows the same image, with the graphene flakes highlighted in red.
c) STM topograph (240\,nm $\times$ 240\,nm) of a full layer of graphene on Ir(111) with a wrinkle. 
The bright line corresponds to the profile given in e).
d) STM topograph (7\,nm $\times$ 7\,nm) of {a wrinkle of low height} in graphene on Pt(111) in atomic resolution. 
The bright line corresponds to the profile given in f).
e) Profile of the topography of the wrinkle in graphene on Ir(111) shown in c). 
f) Profile of the topography of the wrinkle in graphene on Pt(111) shown in d).
STM images have been processed using the WSxM software \cite{horcas2007}.
}
\label{Fig1}
\end{figure}
%
Figure \ref{Fig1}\,a) shows a bright field LEEM image of a graphene flake on Ir(111). The flake has a diameter of $\approx 6\,\mu$m. Branched line defects on the flake which develop upon cooling to room temperature form a network of dark lines, much darker than the substrate step structure which can faintly be seen in the image as well
\cite{fieldeffect}. The STM topograph in figure \ref{Fig1}\,b) shows graphene islands on Pt(111). 
The branched line defects are also present. A typical line defect is shown in figure \ref{Fig1}\,c). It crosses the image diagonally and diverges in two at the bottom of the image. It roughly follows the direction of the two monoatomic steps of the underlying Ir(111) substrate. The defect is about 3\,nm high, and thus much higher than a substrate step as shown in figure \ref{Fig1} e). Its width is a few nanometers as well. On low line defects as shown in figure \ref{Fig1}\,d), profile in figure \ref{Fig1} f), it is possible to achieve atomic resolution on the ridge. The atomic rows over the defect are continuous. The dense packed rows cross the wrinkle roughly perpendicular. The fact that these structures are present only on the graphene flakes and never on the uncovered part of the surface corroborates the assumption that they are indeed wrinkles and not nanotubes on the sample, as has been proposed previously \cite{derycke2002,starr2006, fujita2004,nyapshaev2009}. The continuity of the atomic rows also indicates that the elongated structures are not formed at ruptures where the islands edges roll or bend up. 

\begin{figure}
\includegraphics{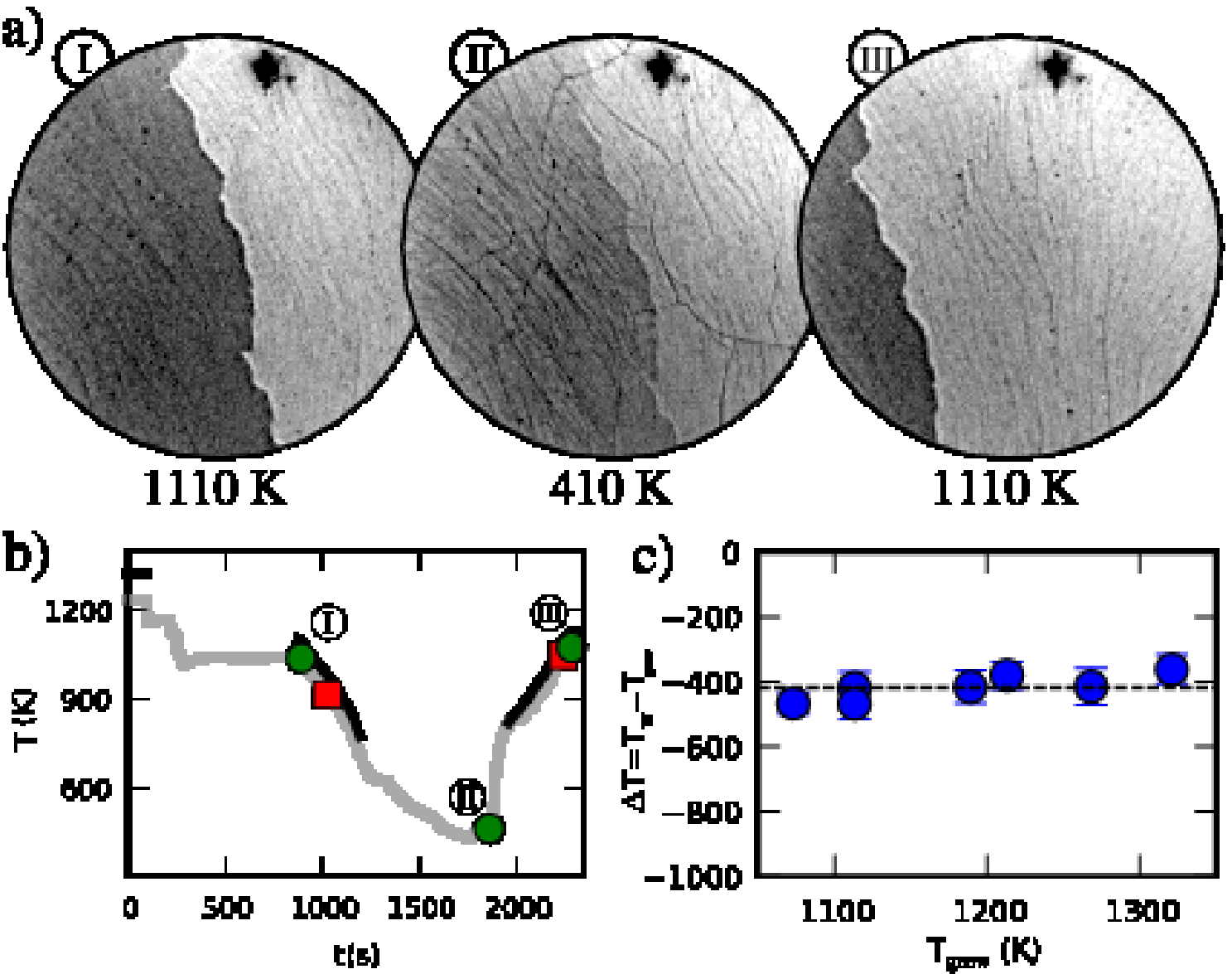}
\caption{
a) Three bright field LEEM images (field of view: 10\,$\mu$m, electron energy: 2.7\,eV) of Ir(111) fully covered with two differently oriented domains of epitaxial graphene. 
Images I and III have been taken at 1110\,K while II has been imaged at 410\,K. 
b) Temperature evolution measured at the sample with a pyrometer (black curve), and at the sample holder with a thermocouple (grey curve).
The green dots mark the points where a) I-III were recorded. 
The red squares mark the onset of wrinkle formation upon cooling and the disappearance of the last wrinkle upon heating.
c)
The difference $\Delta T$ of the onset of wrinkle formation $T_\mathrm{w}$ and the graphene growth temperature $T_\mathrm{grow}$ based on the pyrometer measurement is plotted as a function of $T_\mathrm{grow}$. 
}
\label{Fig2}
\end{figure}
The LEEM images in figure \ref{Fig2}\,a) show epitaxial graphene on Ir(111) fully covering the field of view.
As visible in micro diffraction (see supplement) the darker graphene domain on the left is rotated with respect to the substrate \cite{loginova2009a}.
The left image \lbrack figure \ref{Fig2}\,a)~I\rbrack~has been taken at high temperature (1100\,K), close to the growth temperature ($T_\mathrm{grow}$=1320\,K) of graphene. 
No wrinkles are observed.
During cooling wrinkles appear and spread all over the field of view as visible in figure \ref{Fig2}\,a)~II.
Upon reannealing close to the growth temperature, the wrinkles disappear again \lbrack figure \ref{Fig2}\,a)~III\rbrack.
Faint dark lines due to steps are present at all temperatures.
The time and temperatures the images have been recorded at are marked with green dots in the temperature vs. time diagram in figure \ref{Fig2}\,b).

The appearance of a wrinkle is an instant process within the time resolution of our measurement of 1\,s, while the decay of the wrinkles is a gradual process. Wrinkles decay at slightly higher temperatures than they form. 
The red squares in figure \ref{Fig2}\,b) indicate the formation and decay temperatures.
This also reflects in a hysteresis of the average lattice parameter of graphene as measured by spot profile analysis LEED measurements \cite{tobepublished}.

Graphene has been grown at several temperatures and the onset temperature of wrinkle formation has been recorded.
The onset of wrinkle formation is measured as the occurrence of the first wrinkle in a field of view of 10\,$\mu$m. 
Soon after the occurrence of the first wrinkle additional wrinkles emerge and continue to form even until the sample is cooled down to room temperature.
The first wrinkles appear after a cooldown of 410\,K $\pm$ 40\,K regardless of the growth temperature \lbrack see figure \ref{Fig2}\,c)\rbrack.
Assuming an average difference of the thermal expansion coefficients of Ir and graphene of $7.15\times10^{-6}$ a temperature difference translates to a mismatch in expansion.
During a cooldown by 410\,K iridium shrinks by 0.33\% while a graphene layer only shrinks by 0.03\% \cite{wimper1967, aip_handbook}.
The remaining 0.3\% have to be taken up by compression or wrinkling of the graphene.
This suggests a stress driven process for wrinkle formation.

\subsection{Lattice expansion}
\begin{figure}
\includegraphics{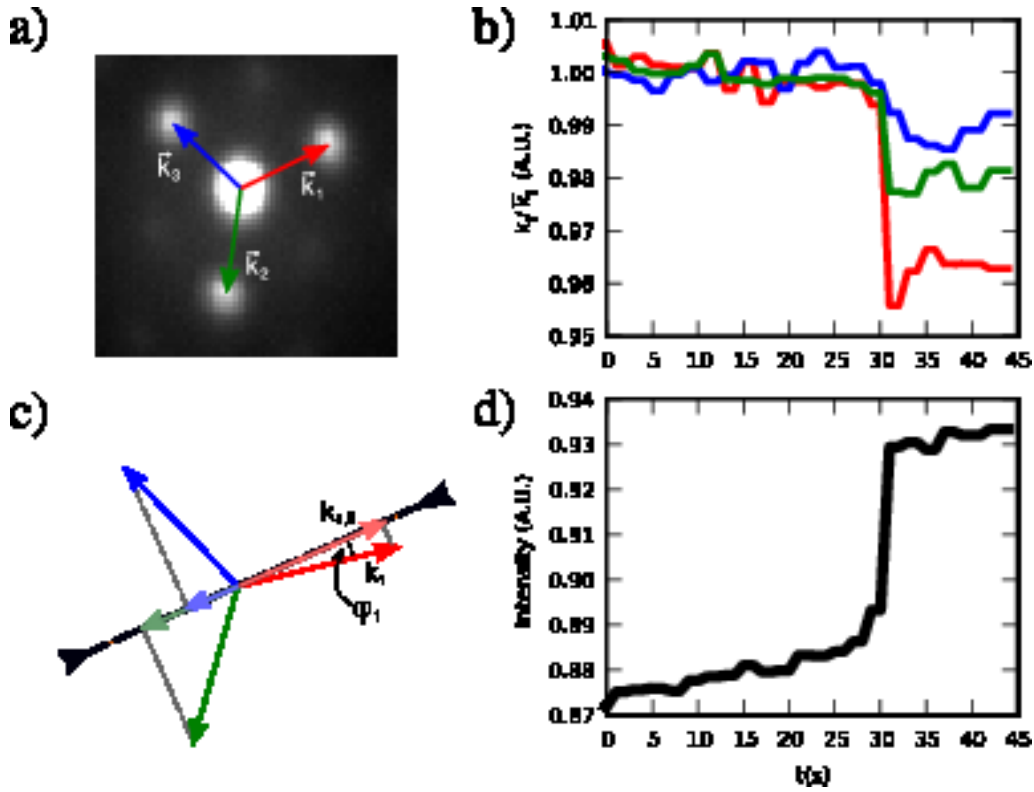}
\caption{
a) Micro diffraction pattern of graphene on Ir(111) near the specular beam with first order moir\'e spots. 
b) Relative length $k_i/\bar{k}_i$ during cooldown (see text).
c) Effect of strain relief in reciprocal space. Only the component of the reciprocal lattice vectors $\vec{k}_i$ which is parallel (light colors) to direction of strain relief (thick black line) is influenced.
d) Intensity of the specular beam during cooldown.
}
\label{Fig3}
\end{figure}
Direct evidence for abrupt strain relief during cooldown is shown in figure \ref{Fig3}. 
Panel a) shows a micro diffraction pattern near the specular beam at 19.9\,eV and resulting from an area
of about 1.4\,$\mu$m diameter. At this energy, three of the six moir\'{e} spots are bright. 
The reciprocal lattice vectors of the moir\'e ${k}_i$ are the difference of the corresponding reciprocal lattice vectors of graphene and Iridium. For small relative changes, their behavior reflects the relative length changes and angular changes in reciprocal lattice of the graphene layer amplified by factor of ten \cite{ndiaye2008}.
The evolution of the relative distances $k_i/\bar{k}_i$ of the bright moir\'{e} spots to the specular beam is displayed by red, green, and blue lines in figure \ref{Fig3} b) \cite{gauss}. Here, $\bar{k}_i$ denotes the time averaged value of $k_i$ during the first 25\,s, i.e. prior to the sudden decrease at about $t=30$\,s.
This decrease amounts to $s_1=(3.6\pm0.4)\%, s_2=(2.0\pm0.4)\%$, and $s_3=(1.2\pm0.5)\%$, where
$s_i = 1 - \bar{k}_i/\bar{k}'_i$ with $\bar{k}'_i$ being the average for the length of the reciprocal moir\'{e} lattice vector between $t=32$\,s and $t=42$\,s. We attribute this decrease to a transition from the flat state into the wrinkled state (see section 3.3). 
The wrinkled graphene is relaxed and has a larger lattice parameter and thus shorter reciprocal lattice vectors
$\vec{k}_i$ \cite{rot_also}. The three $\vec{k}_{i}$ shrink by different amounts.
This is well understandable given the wrinkle is a linear defect which can only relax stress in one direction.
Assuming all three $\vec{k}_i$ to be of equal length and making angles of exactly 120$^\circ$ prior to wrinkling, we can estimate the direction of stress relaxation.
Each vector $\vec{k}_{i}$ can be split into a component which is parallel to the direction of strain relief and one which is perpendicular to this direction.
Only the parallel component is affected by the strain relief. Let $\phi_i$ be the angle between the lattice vector $\vec{k}_i$ and the direction of strain relief, $\bar{k}_i$ and {$\bar{k}'_i$} the averaged lengths of the reciprocal lattice vectors before and after the transition, $\bar{k}_{i,\parallel} =\bar{k}_i \cos\phi_i$ the average component parallel and $\bar{k}_{i,\bot} = \bar{k}_i \sin\phi_i$ the average component of $\vec{k}_i$ 
perpendicular to the direction of strain relaxation and prior to the relaxation and finally $c = \bar{k}'_{i,\parallel}/\bar{k}_{i,\parallel}$ the reciprocal space compression upon wrinkling. 
This leads to  
\[1- \left(\frac{\bar{k}'_i}{\bar{k}_i}\right)^2 = (1-c^2)\cos^2\phi_i\]
for all three reciprocal linearly dependent lattice vectors.
With the measured values for $\frac{\bar{k}'_{i}}{\bar{k}_{i}}$ this equation can be numerically solved. 
This leads to a factor $c=96\%$ corresponding to a strain relief of 0.4\% in the graphene layer.
The angle between the direction of compression and {$\vec{k}_1$} is roughly $\phi_1=10^\circ$  as illustrated in figure \ref{Fig3} c). This implies that the wrinkling took place roughly perpendicular to the dense packed rows of the graphene layer consistent with what is expected from STM data as the dominant orientation of wrinkles (cf. figure \ref{Fig1} d) or \cite{sun2009})
The intensity of the specular beam abruptly increases simultaneously with the relaxation of the lattice as shown in figure \ref{Fig3} d).

\subsection{Local stress evolution}
\begin{figure}
\includegraphics{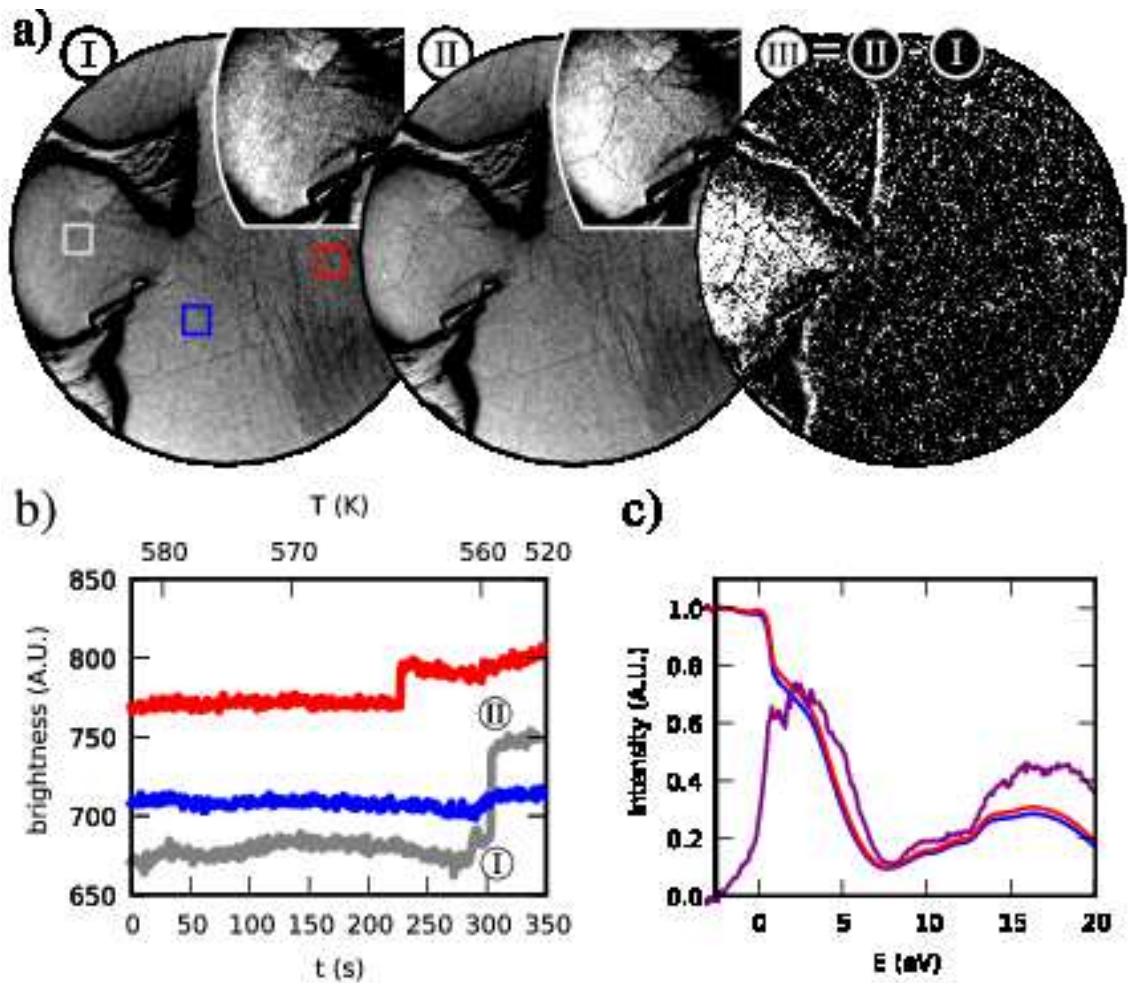}
\caption{
a) Two subsequent LEEM images of graphene on Ir(111) recorded during cooling (field of view: 10\,$\mu$m, electron energy: 2.5\,eV) and the difference of their intensities with enhanced contrast.
Between (I) and (II) a wrinkle forms on the peninsula on the left. This is shown enlarged and contrast enhanced in the insets.
The difference image shows that the intensity has increased locally in the course of wrinkle formation. 
b) The intensity integrated over the regions marked by colored boxes in a)\,I over time, with linear background subtraction.
Whenever a wrinkle is created, the brightness increases abruptly. 
The roman numbering indicates where a)\,I and a)\,II have been recorded.
c) LEEM I(V)-curve. Blue: unwrinkled graphene, red: wrinkled graphene, purple: difference ($\times 20$).
}
\label{Fig4}
\end{figure}

The abrupt intensity change of the specular beem upon wrinkling allows us to monitor the local extension of stress relaxation through LEEM imaging. Figure \ref{Fig4}\,a) shows LEEM images of graphene on Ir(111). The sample is partially covered by graphene prepared at 1110\,K and the sample has been cooled down to 560\,K within one hour. In the course of this cooling, some wrinkles have already formed, especially in the right hand side of the field of view. The images I and II have been measured subsequently with a delay of 1\,s.
They capture a single event of wrinkle formation on the graphene patch near the left border of the image. 
Simultaneously with wrinkle formation the brightness increases in the affected area.
This change is visualized in figure \ref{Fig4}\,a) III which is the difference between the images in II and I. 
It shows that the formation of a wrinkle does not only act nanoscopically at the line of delamination but it rather has an impact on a mesoscopic scale (in this case 4\,$\mu \mathrm{m}^2$).

The change in intensity integrated over the regions marked in figure \ref{Fig4}\,a)~I is shown in figure \ref{Fig4}\,b). This change takes place at different times, locations and to different extents. We interpret the increase in image intensity as an effect of the relaxation of the graphene lattice. 
This provides an explanation for the observation of locally varying compression in epitaxial graphene \cite{robinson2009}.

To discuss the origin of the $(0,0)$-spot intensity increase upon wrinkling we consider figure \ref{Fig4}\,c). It shows I(V)-curves of the $(0,0)$ spot of flat and wrinkled graphene on Ir(111) and their difference. The curves were taken from one sequence of images recorded at intermediate temperatures where wrinkles have just started to form and some parts of the surface are still unwrinkled. The difference curve has two maxima. Both are correlated with maxima in total intensity resulting from 
constructive interference of electron waves reflected from graphene and the Ir surface. For a distance of 3.4\,\AA~\cite{feibelman2008} we expect maxima at 3.2\,eV and 13\,eV. 
Therefore we speculate that the intensity changes are due to structural changes, such as a change in the spacing between the graphene layer and the Ir substrate triggered by the relaxation of the graphene lattice. 
Evidently these structural relaxations are accompanied by changes in the graphene electronic structure
which additionally may affect the I(V)-curve.
 

During a wrinkling event many atoms are displaced.
On Ir(111) and Pt(111) graphene forms an incommensurate superstructure \cite{ndiaye2008, land1992}.
That implies a low barrier for sliding of graphene on the surface, because for every atom, which loses energy by moving out of its optimum binding configuration another atom gains energy.
The small flakes grown by temperature programmed growth \cite{coraux2009} barely show wrinkles. 
A graphene flake smaller than the average separation of two wrinkles just expands if the compression gets large enough to overcome the barrier for sliding. 
For smaller islands, this barrier may be larger, due to edge effects becoming more relevant.
Accordingly there is residual strain in such flakes \cite{ndiaye2008}.

Wrinkle patterns from repeated cooling and heating cycles at one sample spot are similar.
This suggests that wrinkles nucleate at preexisting features. 
We find in our STM data spots of delamination which are centered at heptagon-pentagon pairs of carbon rings (see supplement) \cite{coraux2008}.
This seems reasonable as these defects induce additional local stress into the graphene lattice. 
We thus speculate that heptagon-pentagon defects are sites of wrinkle nucleation.

\subsection{Model}
\begin{figure}
\includegraphics{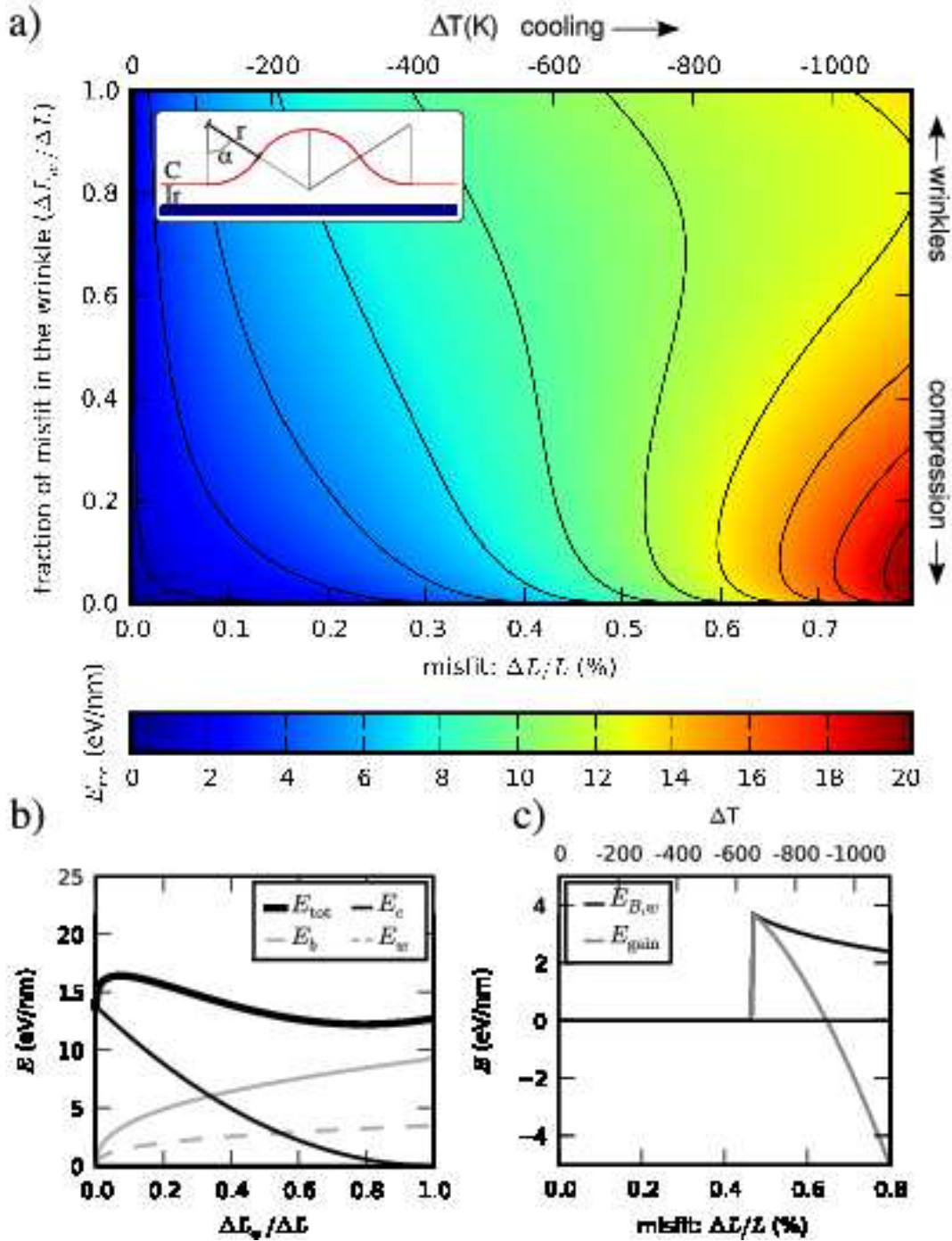}
\caption{
a) A map of the energy per nm of a wrinkle $E_\mathrm{tot}$ with respect to uncompressed flat graphene according to a one-dimensional continuum model. 
The shape of the wrinkle has been modeled as four arcs of circles (see inset). 
The lower horizontal axis shows the misfit of graphene and Ir $\frac{\Delta L}{L}$, the upper axis shows the corresponding temperature difference $\Delta T$.
The vertical axis represents the fraction of the misfit which is compensated by wrinkling $\frac{\Delta L_w}{\Delta L}$ instead of compression.
b) The graph shows the energy of a wrinkle per nm for a mismatch of 0.7\% (thick black line) and the contributions it consists of. 
There is a local minimum for the flat configuration where all the energy is stored in the form of compression $E_c$, while in the global energy minimum most of the energy is stored in a wrinkle as bending and reduced bonding.
c) Energy barrier $E_{B,w}$ for wrinkle formation. 
The system can gain energy by wrinkle formation ($E_\mathrm{gain}$) with respect to compressed flat graphene, if the misfit is larger than 0.65\%.
}
\label{Fig5}
\end{figure}
The wrinkle formation can be described in a one-dimensional continuum model.
When the substrate and graphene cool down, the graphene has to compensate for the thermal misfit $\Delta L/L$ resulting from the difference of the thermal expansion coefficients. 
Either compression $\Delta L_c/L$ or the formation of a wrinkle $\Delta L_w/L$ can compensate for that misfit ($\Delta L = \Delta L_c + \Delta L_w$).

To calculate the energy of compression we use the separation between two wrinkles $L=260$\,nm as estimated from experiment.
With an atom density of 
$n=36.2$\,atoms/nm$^{2}$
and an elastic modulus of $Y=56$\,eV per atom \cite{jiang2008} the compression energy per nm of wrinkle length can be expressed as $E_c = \frac{1}{2}(\Delta L_c/L)^2 L\cdot Y\cdot n$. 
The energy necessary for wrinkle formation consists of two contributions: 
first, there is the bending of the graphene layer. 
An estimation for this contribution is available from the study of single walled carbon nanotubes \cite{jiang2008},
giving the bending energy per atom in a nanotube of radius $r$ as $e_w=a/r^2+b/r^4$ with the empirical parameters $a=0.99$\,eV/\AA$^2$ and $b=12.3$\,eV/\AA$^4$.  
Second, bonds between graphene and the substrate are stretched or broken, where the graphene flake delaminates. 
Since the estimated height of graphene on Ir(111) is comparable to the interlayer distance of graphite, we use the exfoliation energy of graphite as an assessment for the binding energy of an atom in a graphene layer on Ir(111) in this case. It has the value of $e_{b,0}=0.052eV$ \cite{zacharia2004, dftbond}
The strength of the van-der-Waals binding energy between a particle and a surface decreases with the cube of the distance, so we calculate the change in binding energy of an atom in an elevated part of the graphene sheet with the height of $z$ instead of $z_0$ above the substrate as $e_b=e_{b,0}\left(1-z_0^3/z^3\right)$.

With $E_w$ and $E_b$ as the integral of $e_w$ and $e_b$ over the atoms in a nm of wrinkle the resulting energy cost for a nm of wrinkle compared to flat relaxed graphene sums up to
\begin{eqnarray*}
E_\mathrm{tot} &=& E_c + E_w + E_b  \\
&=& \frac{1}{2}(\Delta L_c/L)^2 \cdot Y\cdot L \cdot n \\
&+& n\int_0^{L-\Delta L_c} \left( \frac{a}{r(x)^2} + \frac{b}{r(x)^4}   \right)dx\\
&+& n\int_0^{L-\Delta L_c} \left( e_{b,0}\left(1-\frac{z_0^3}{z^3}\right) \right)dx\
\end{eqnarray*}
and leads to a complex variation problem for the shape of the wrinkle.

Here we assume a simple shape for the wrinkle, consisting of four equal arcs of a circle with radius $r$ and an opening angle $\alpha$ as shown in the inset of figure \ref{Fig5}\,a). 
The model contains $r$ (or $\alpha$) as a free parameter which is optimized for minimum energy $E_{tot}$ for each combination of $\Delta L / L$ and $\Delta L_w / L$.

In figure \ref{Fig5}\,a) $E_\mathrm{tot}$  is plotted for this model as a function of thermal misfit $\Delta L / L$ (lower horizontal axis) and fraction of strain accommodated in a wrinkle $\Delta L_w/\Delta L$ (vertical axis). 
Moving along the horizontal axis from left to right corresponds to cooling of the sample.
The according temperature difference is indicated on the upper horizontal axis. 
In the lower part of the diagram most of the misfit is taken up by compression while in the upper part the misfit is compensated for by a wrinkle.
For misfits below 0.3\% there is exactly one optimum configuration: The graphene layer is compressed and there is no wrinkle.
As the misfit increases (temperature decreases), a second local minimum in energy emerges. 
Nevertheless, the unwrinkled compressed state still is favorable. 
For $\Delta L/L > 0.65$\% of misfit, a situation rendering about 80\% of the misfit subject to wrinkle formation, is optimal.

A cut through the map at constant misfit $\Delta L/L=0.7\%$ is shown in figure \ref{Fig5}\,b). 
There is a local minimum for the flat configuration where all the energy is stored in the form of compression $E_c$, but the optimum configuration is the formation of a wrinkle, which contains most of the energy in the form of reduced bonding to the substrate ($E_b$) and bending of the graphene ($E_w$). 
Figure \ref{Fig5}\,c) illustrates the relationship of the two energy minima and the barrier in between.
The gray line shows the difference in energy of the unwrinkled state and the wrinkled state $E_\mathrm{gain}$.
For a misfit below 0.47\% there is no minimum for the wrinkled state, above 0.47\%, there is a local minimum, but its energy is higher than that of the uncompressed flat state.
Only for compressions above 0.65\%, when the gray line enters the negative region, the system can gain energy by forming a wrinkle.
Still, there is an energy barrier to overcome, which allows the system to be trapped in the local minimum explaining the sudden and abrupt formation of wrinkles.
This is consistent with the hysteresis for wrinkle appearance and disappearance.

Although our one-dimensional model explains all qualitative features observed, the model prediction overestimates the experimentally observed critical misfit for wrinkling formation nearly by a factor of two.
Certainly a full two dimensional analysis may lead to somewhat different numbers -- wrinkle formation is likely to be eased by biaxially compressed graphene. Also the wrinkle separation $L$, the binding energy $E_{b,0}$ and our simple model for the wrinkle shape carry significant uncertainties.

As wrinkles are large scale defects, it would be desirable to suppress their formation.
One way to achieve this could be to reduce the amount of total thermal misfit by growing graphene at the lowest possible temperature and inserting an intermediate annealing step to remove the defects prior to cooldown. Combining temperature programmed growth and chemical vapor deposition it appears possible to achieve high quality graphene at a growth temperature of only 1000\,K \cite{Gastel2009}.  
Also grazing incidence keV ion erosion removing exclusively protruding wrinkles followed by annealing could lead to continuous graphene with less or no wrinkles.
A third approach could be to increase the energy for bending. This could be accomplished by evaporating at high substrate temperatures a thin film with a thermal expansion coefficient similar to the substrate on top of the graphene. This cover layer would have to be bent as well, for the graphene to form a wrinkle.

\section{Conclusion}
In conclusion, we demonstrated how LEEM can be used to monitor strain relaxation in situ. It was possible to develop a consistent picture of wrinkle formation on graphene linking wrinkle formation with inhomogeneous residual strain. The development of wrinkles appears to be a serious problem for all methods of growth of weakly bound epitaxial graphene, as all require high temperatures. We hope the improved understanding of wrinkle formation achieved here will contribute to a future solution of this problem.

\section{Acknowledgement}
Financial support by Spain's MEC under grant No. MAT2007-60686 
and Deutsche Forschungsgemeinschaft is gratefully acknowledged. 
J. C. was supported by a Humboldt fellowship.

\section{Supporting information}
S1: LEEM, photo electron emission microscopy and microdiffraction measurements of rotational domains,  STM data of a delaminated bulge around a dislocation.
S2: Movie with brightness increase upon wrinkle formation as in figure \ref{Fig4}.

\bibliographystyle{unsrt}
\bibliography{graphene-wrinkles}

\begin{thebibliography}{10}
\bibitem{fieldeffect}
The rim of the flake appears very broad and dark, because the electric field at
  the edge of the island deteriorates the imaging electron beam.
\bibitem{gauss}
The distances of the moir\'{e} spots from the (0,0) spot are determined by the
  difference\tm{s} of the reciprocal lattice vectors of graphene and Ir(111)
  and amount to about 20 pixels in the original data. To determine these
  distances with sub pixel resolution, a two dimensional gauss peak has been
  fitted into each LEED-spot. The centers of the gauss peaks accurately point
  to the center of the spots.
\bibitem{rot_also}
Slightly rotated domains can also lead to changes in the moir\'e repeat
  distance. This is excluded here.
\end{thebibliography}

\end{document}


\title{In situ observation of stress relaxation in epitaxial graphene: Supplement I/II}
\author{
Alpha T. N'Diaye$^1$\footnote{Corresponding autor; Email: ndiaye@ph2.uni-koeln.de},
Raoul van Gastel$^2$,
Antonio J. Mart\'inez-Galera$^3$,
Johann Coraux$^1$\footnote{Permanent address: Institut N{\'e}el/CNRS-UJF, 25 rue des Martyrs, BP 166, 38042 Grenoble cedex 9, France}, 
Hichem Hattab$^4$,
Dirk Wall$^4$,
Frank-J. Meyer zu Heringdorf$^4$,
Michael Horn--von Hoegen$^4$,
Jos\'e M. G\'omez-Rodr\'iguez$^3$,
Bene Poelsema$^2$,
Carsten Busse$^1$,
Thomas Michely$^1$
}

\section{Parallel and rotated graphene phases in LEEM and PEEM}

\begin{figure}
\includegraphics{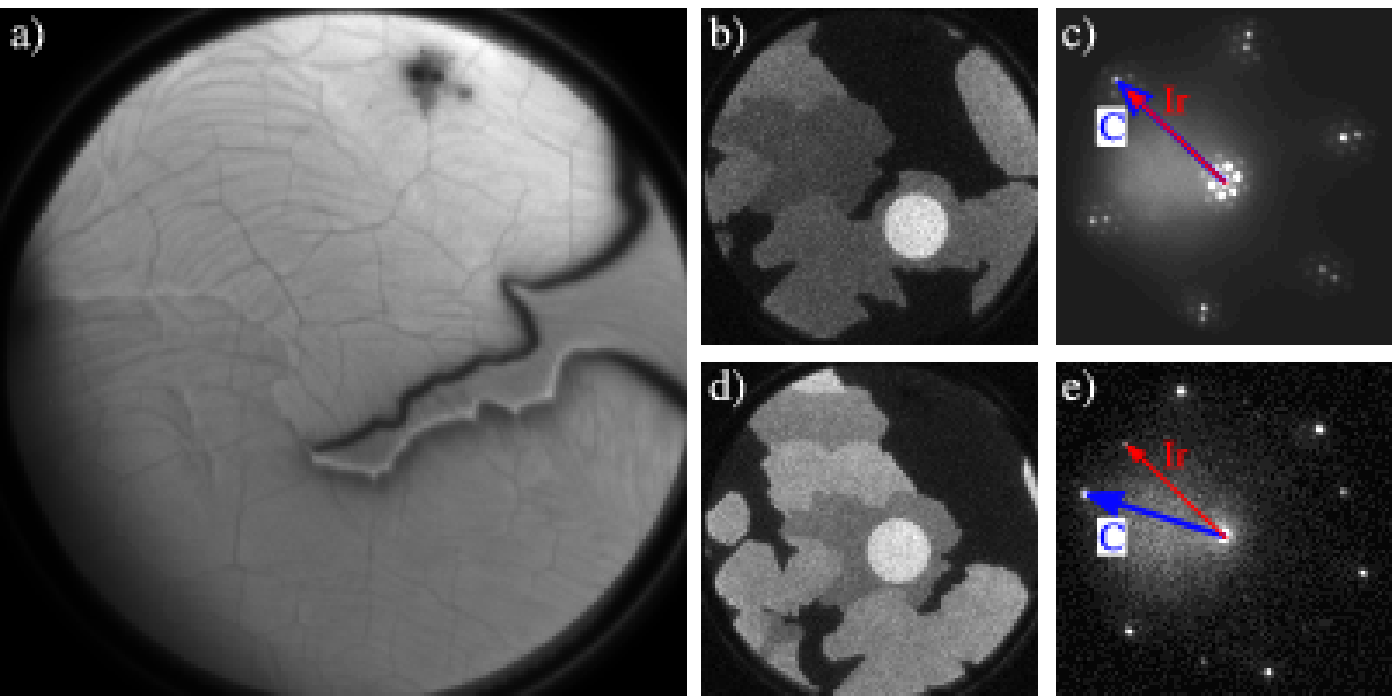}
\caption{
a) bright field LEEM image (field of view: 15\,$\mu$m, e$^-$--energy: 2.68\,eV) 
of the two graphene phases and uncovered Ir(111) at room temperature.
b) Zoomed out PEEM image of the same area as in a) (field of view: 25\,$\mu$m)
The area used for the microdiffraction pattern in d) is illuminated by the electron beam (bright circle).
c) microdiffraction pattern of the parallel graphene phase (electron energy: 45.9\,eV)
d) Zoomed out PEEM image of the same area as in a) (field of view: 25\,$\mu$m)
The bright spot marks the area used for the microdiffraction pattern in e)
e) microdiffraction pattern of the rotated graphene phase (electron energy: 45.9\,eV)
}
\label{Fig1}
\end{figure}
We observe two domains of graphene on Ir(111). 
In \ref{Fig1} a) a bright field LEEM image is shown.
Graphene in the upper area is imaged brighter than graphene in the lower area, and there is a gulf of bare iridium cutting in from the right.
The images b) and e) are photo emission electron microscopy (PEEM) images of the same region. 
Iridium is imaged black, graphene from the upper area of a) is imaged darker than graphene from the lower area.
The bright circle displays the position of the electron beam for the LEED images in c) and e).
The LEED patterns show that in the case of bright graphene in LEEM (dark in PEEM), there is a rotation of graphene's $[11\bar{2}0]$ direction with respect to the substrate's $[10\bar{1}]$ direction whereas in the dominating phase - darker in LEEM (bright in PEEM) the dense packed directions of graphene and Ir(111) are parallel. 

\section{Delamination around heptagon-pentagon pairs}
\begin{figure}[h!]
\includegraphics{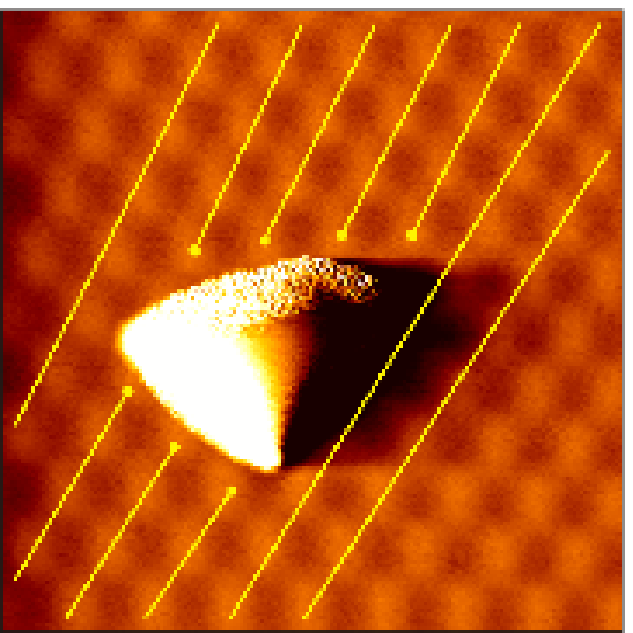}
\caption{
STM topograph of graphene on Ir(111) (20\,nm $\times$ 20\,nm). The moir\'e rows around the bulge are marked with bright lines.
}
\label{Fig2}
\end{figure}
The STM topograph in \ref{Fig2} shows a bulge in the graphene layer. 
The extra ending line of moir\'e maxima (three on one side of the protrusion, but four on the other) indicate an extra row of atoms terminating in a pair of heptagon-pentagon carbon rings \cite{coraux2008}.
We speculate that the bulge is a nucleus of a wrinkle originating from such a point defect.

\bibliographystyle{unsrt}
\bibliography{graphene-wrinkles}